\documentclass[aps,prb,superscriptaddress, longbibliography, reprint]{revtex4-2}
\makeindex
\usepackage{amsmath}  	
\usepackage{graphicx}  	
\usepackage{verbatim}  	
\usepackage{color}   	
\usepackage{subfigure} 	
\usepackage{hyperref}  
\usepackage{gensymb}
\usepackage{xcolor}
\usepackage[dvipsnames]{xcolor}
\usepackage{nicefrac,xfrac}
\usepackage{soul}
\byrevtex

\hypersetup{
 colorlinks=true,
 linkcolor=blue,
 citecolor=blue,
 urlcolor=blue
}

\begin{document}
\title{Evolution of charge-density-wave soft phonon modes in $\mathrm{Pd}_x\mathrm{ErTe}_3$}

\affiliation{Heinz Maier-Leibnitz Zentrum (MLZ), Technische Universität München, D-85747 Garching, Germany}
\affiliation{Materials Science Division, Argonne National Laboratory, Lemont, Illinois 60439, USA}
\affiliation{Institute for Quantum Materials and Technologies, Karlsruhe Institute of Technology, Kaiserstr. 12, D-76131 Karlsruhe, Germany}
\affiliation{Advanced Photon Source, Argonne National Laboratory, Lemont, Illinois 60439, USA}
\affiliation{Stanford Institute for Materials and Energy Sciences, SLAC National Accelerator Laboratory, 2575 Sand Hill Road, Menlo Park, California 94025, USA}
\affiliation{Geballe Laboratory for Advanced Materials, Stanford University, Stanford, California 94305, USA}
\affiliation{Department of Applied Physics, Stanford University, Stanford, California 94305, USA}

\author{Avishek Maity}
\email{Contact author: maitya@ornl.gov}
\thanks{Present address: Neutron Scattering Division, Oak Ridge National Laboratory, Oak Ridge, Tennessee 37831, USA}
\affiliation{Heinz Maier-Leibnitz Zentrum (MLZ), Technische Universität München, D-85747 Garching, Germany}

\author{Stephan Rosenkranz}
\affiliation{Materials Science Division, Argonne National Laboratory, Lemont, Illinois 60439, USA}

\author{Raymond Osborn}
\affiliation{Materials Science Division, Argonne National Laboratory, Lemont, Illinois 60439, USA}

\author{Rolf Heid}
\affiliation{Institute for Quantum Materials and Technologies, Karlsruhe Institute of Technology, Kaiserstr. 12, D-76131 Karlsruhe, Germany}

\author{Ayman H. Said}
\affiliation{Advanced Photon Source, Argonne National Laboratory, Lemont, Illinois 60439, USA}

\author{Ahmet Alatas}
\affiliation{Advanced Photon Source, Argonne National Laboratory, Lemont, Illinois 60439, USA}

\author{Joshua A. W. Straquadine}
\affiliation{Stanford Institute for Materials and Energy Sciences, SLAC National Accelerator Laboratory, 2575 Sand Hill Road, Menlo Park, California 94025, USA}
\affiliation{Geballe Laboratory for Advanced Materials, Stanford University, Stanford, California 94305, USA}
\affiliation{Department of Applied Physics, Stanford University, Stanford, California 94305, USA}

\author{Matthew J. Krogstad}
\affiliation{Advanced Photon Source, Argonne National Laboratory, Lemont, Illinois 60439, USA}

\author{Anisha G. Singh}
\affiliation{Stanford Institute for Materials and Energy Sciences, SLAC National Accelerator Laboratory, 2575 Sand Hill Road, Menlo Park, California 94025, USA}
\affiliation{Geballe Laboratory for Advanced Materials, Stanford University, Stanford, California 94305, USA}
\affiliation{Department of Applied Physics, Stanford University, Stanford, California 94305, USA}

\author{Ian R. Fisher}
\affiliation{Stanford Institute for Materials and Energy Sciences, SLAC National Accelerator Laboratory, 2575 Sand Hill Road, Menlo Park, California 94025, USA}
\affiliation{Geballe Laboratory for Advanced Materials, Stanford University, Stanford, California 94305, USA}
\affiliation{Department of Applied Physics, Stanford University, Stanford, California 94305, USA}

\author{Frank Weber}
\email{Contact author: frank.weber@kit.edu}
\affiliation{Institute for Quantum Materials and Technologies, Karlsruhe Institute of Technology, Kaiserstr. 12, D-76131 Karlsruhe, Germany}

\begin{abstract}

We investigated the lattice dynamics of quasi-two-dimensional Pd-intercalated $\mathrm{ErTe}_3$ in relation to its charge-density-wave (CDW) transitions by means of x-ray diffuse and meV-resolution inelastic x-ray scattering. In pristine $\mathrm{ErTe}_3$, CDW order develops at orthogonal in-plane wave vectors $\boldsymbol{\mathrm{q}}_{1}^{c} = (0, 0, 0.29)$ (the $c\text{-}\mathrm{CDW}$) and $\boldsymbol{\mathrm{q}}_{2}^{a} = (0.31, 0, 0)$ (the $a\text{-}\mathrm{CDW}$), with transition temperatures $T_{1}^{c} = 270$~K and $T_{2}^{a} = 160$~K, respectively. Remarkably, we observe diffuse x-ray scattering already near the higher transition temperature $T_{1}^{c}$ along $a\text{-}\mathrm{CDW}$ but at a slightly different wave vector $\boldsymbol{\mathrm{q}}_{1}^{a} = (0.29, 0, 0)$. Inelastic x-ray scattering for $\mathrm{Pd}_{0.01}\mathrm{ErTe}_3$ shows that a partial phonon softening at $\boldsymbol{\mathrm{q}}_{1}^{a}$, underscoring the strong competition between ordering tendencies along the nearly equivalent in-plane axes of the orthorhombic lattice. For intercalation levels $x \geq 0.02$, the $a\text{-}\mathrm{CDW}$ state is suppressed. Nevertheless, a similar correlation between phonon softening and diffuse scattering persists along the $[100]$ direction, again observed at $\boldsymbol{\mathrm{q}}_{1}^{a} = (0.29, 0, 0)$ and $T_{1}^{c}$. These findings suggest that the $a\text{-}\mathrm{CDW}$ is fully suppressed for $x \geq 0.02$, and that the residual diffuse scattering at $\boldsymbol{\mathrm{q}}_{1}^{a}$ originates from the partial phonon softening associated with the $c\text{-}\mathrm{CDW}$, reflected by the near equality of the absolute size of $\boldsymbol{\mathrm{q}}_{1}^{c}$ and $\boldsymbol{\mathrm{q}}_{1}^{a}$.

\end{abstract}

\maketitle

\section{INTRODUCTION}

Disorder plays a decisive role in destabilizing \textcolor{blue}{charge-density-wave (CDW)} order by disrupting the long-range periodic modulation of the lattice and electronic density. In the rare-earth tritelluride $\mathrm{ErTe}_3$, Pd intercalation has emerged as a prototypical route to introduce controlled quenched disorder, allowing systematic studies of its impact on competing unidirectional CDWs: Diffraction and \textcolor{blue}{scanning tunneling microscopy (STM)} measurements reveal that increasing Pd concentration broadens and weakens the superlattice reflections, reflecting shortened CDW coherence lengths and the proliferation of dislocations in the modulation pattern \citep{Straquadine_prb_2019,Fang_prb_2019}. More recent x-ray studies have identified signatures of a Bragg-glass–like state, where residual quasi-long-range order persists despite significant disorder \citep{Mallayya_nphys_2024}. Complementary electrodynamic and cryo-STEM experiments further demonstrate that defects fill in the CDW gap and can realign or locally suppress modulations \citep{Corasaniti_prr_2023,Siddique_prb_2024}. Together, these findings highlight how quenched disorder progressively degrades CDW order in $\mathrm{ErTe}_3$, transforming sharp superlattice peaks into diffuse scattering features and thereby providing a direct experimental window into the interplay between collective electronic order and defects in low-dimensional materials which we will further investigate by \textcolor{blue}{x-ray diffuse scattering (XDS)} and meV-resolution \textcolor{blue}{inelastic x-ray scattering (IXS)}.

The rare-earth tritelluride RTe$_3$ compounds, spanning almost the entire rare-earth series from R = La-Nd, Sm, to Gd-Tm \citep{Bucher_prb_1975,Norling_inorgchem_1966,DiMasi_prb_1995}, provide a unique opportunity to explore the physics of CDW formation in a model system for unidirectional CDWs in quasi-two-dimensional metals with an almost square lattice \citep{DiMasi_prb_1995, Gweon_prl_1998}. Recent research provided evidence for intriguing properties such as a light-induced competition of different CDW states \citep{Kogar_nphys_2020} and an axial Higgs mode \citep{Wang_nature_2022}. Early work focused on the fundamental properties of pristine $\mathrm{ErTe}_3$ \citep{Ru_prb_2008,Ru_prb_2008_b} whereas more recent publications discuss the evolution of CDW states upon intercalation \citep{Mallayya_nphys_2024,Straquadine_prb_2019,Fang_prb_2019}, strain application \citep{Straquadine_prx_2022, Singh_sciadv_2024,Singh_prb_2025} or after light excitations \citep{Su_arXiv_2025}. 

The RTe$_3$ exhibit a room temperature orthorhombic structure ($\mathrm{ErTe}_3$: $Cmcm$; $a_\mathrm{RT} = 4.28432$ \AA, $b_\mathrm{RT} = 25.24484$ \AA, $c_\mathrm{RT} = 4.29057$ \AA, $c_\mathrm{RT}/a_\mathrm{RT} \approx 1.001$). The orthorhombic distortion arises from a glide plane along the in-plane $c$-axis, which governs the stacking of the $\mathrm{RTe}$ layers and causes a slight difference in the two in-plane lattice constants. For all R = La-Nd, Sm, to Gd-Tm, the materials primarily undergo a unidirectional incommensurate $c\text{-}\mathrm{CDW}$ ordering along the in-plane $c$-axis \citep{DiMasi_prb_1995, Ru_prb_2006, Ru_prb_2008, Ru_prb_2008_a}. The corresponding transition temperature $T_{1}^{c}$ is influenced by the size of the $\mathrm{R}$ elements and varies from above 500 K in $\mathrm{LaTe}_3$ to 250 K in $\mathrm{TmTe}_3$. Additionally, the materials with heavier $\mathrm{R}$ elements (Tb, Dy, Ho, Er, and Tm) feature a second unidirectional CDW with a lower transition temperature $T_{2}^{a}$ ranging from 41 K in $\mathrm{TbTe}_3$ to 180 K in $\mathrm{TmTe}_3$ in the perpendicular direction to the first one, i.e. along the in-plane $a$-axis \citep{Ru_prb_2008_b,Hu_prb_2011,Banerjee_prb_2013,Hu_prb_2014}.

Several studies have suggested that these two CDW orders compete with each other because suppression of $T_{1}^{c}$ enhances $T_{2}^{a}$, e.g. under chemical \citep{Banerjee_prb_2013} and hydrostatic pressure \citep{Sacchetti_prl_2007,Lavagnini_prb_2009,Hamlin_prl_2009}. On the other hand, disorder by intercalation suppresses both CDW transitions simultaneously e.g. in $\mathrm{Pd}_x\mathrm{ErTe}_3$ \citep{Straquadine_prb_2019}. Investigation of lattice dynamics on $\mathrm{DyTe}_3$ using IXS \citep{Maschek_prb_2018} found a classic phonon softening only at $T_{1}^{c}$ and $\boldsymbol{\mathrm{q}}_{1}^{c}$ but not for $T_{2}^{a}$ and $\boldsymbol{\mathrm{q}}_{2}^{a}$ indicating that the two different charge orders are not simply a $90^{\circ}$-rotated version of each other.

Here, we report XDS measurements in $\mathrm{Pd}_x\mathrm{ErTe}_3$, $x = 0$ - $0.029$, and meV-resolution IXS for samples with $x = 0.01$ and $0.023$ to track the evolution of the CDW soft phonon modes as CDW orders are suppressed in Pd-intercalated $\mathrm{ErTe}_3$ \citep{Straquadine_prb_2019}. Our main findings are that XDS along the $a^*$ direction peaks at the $c\text{-}\mathrm{CDW}$ transition temperature. The long-range ordered $a\text{-}\mathrm{CDW}$ is suppressed around $x = 0.01$ and its superlattice peaks are replaced by XDS at higher intercalation levels indicating short-to-medium range order only. Importantly, the XDS along $a^*$ peaks at the respective $c\text{-}\mathrm{CDW}$ transition temperature, $T_{1}^{c}$, and is observed at a wave vector $\boldsymbol{\mathrm{q}}_{1}^{a}$ which is slightly different from that of the long-range ordered $a\text{-}\mathrm{CDW}$, $\boldsymbol{\mathrm{q}}_{2}^{a}$, but close in absolute value to the ordering wave vector of the $c\text{-}\mathrm{CDW}$, $\boldsymbol{\mathrm{q}}_{1}^{c}$. 
Thus, the combined analysis of XDS and phonon spectroscopy indicates that the XDS along $a^*$ in $\mathrm{ErTe}_3$ at $T>T_{2}^{a}$ and in $\mathrm{Pd}_x\mathrm{ErTe}_3$ at $x > 0.01$ is related to a phonon softening strongest at $T_{1}^{c}$. Interestingly, we find no phonon softening at $T_{2}^{a}$ which, taken together with recent reports \citep{Maschek_prb_2018,Su_arXiv_2025}, suggests a different mechanism underlying the two CDW orders present in $\mathrm{Pd}_x\mathrm{ErTe}_3$.

\section{METHODS}

XDS measurements were performed at sector 6-ID-D at the Advanced Photon Source, Argonne National Laboratory \citep{Krogstad_nmat_2020}. Samples were mounted on the tips of polyimide capillaries and cooled using an Oxford N-Helix Cryostream, which surrounded samples with either N$_2$ or He gas. Measurements were taken with incident x-ray energy of 87 keV in transmission geometry, with samples continuously rotated at $1^{\circ}$ per second and a Pilatus 2M CdTe detector taking images at 10 Hz. NXRefine module of the NeXpy software was used to analyze the XDS data \citep{nxrefine_github_2024}. More detailed information are provided in \citep{Mallayya_nphys_2024,Singh_prb_2025} where part of the XDS data and corresponding analysis have been published.

The IXS experiments were performed at the 30-ID high energy-resolution IXS (HERIX) beamline of the Advanced Photon Source, Argonne National Laboratory \citep{Toellner_jsr_2011,Said_jsr_2011}, with an incident energy of 23.78 keV and a resolution of about 1.5 meV. Phonon, i.e., energy scans were done at constant momentum transfer and the components of the scattering vector $\boldsymbol{\mathrm{Q}}$ are expressed in reciprocal lattice units (r.l.u) as $(Q_h,Q_k,Q_l)= (2\pi/a, 2\pi/b, 2\pi/c)$ ($a$, $b$, $c$: lattice constants). Phonon excitations in the the constant-$\boldsymbol{\mathrm{Q}}$ scans were approximated using damped harmonic oscillator (DHO) functions \citep{Fak_physicab_1997}, convoluted with the experimental resolution function obtained by scanning a piece of plastic. The DHO function is defined as
\[
S(\mathbf{Q},\omega)=
\frac{[n(\omega)+1]\,Z(\mathbf{Q})\,4\omega\Gamma/\pi}
{(\omega^2-\tilde{\omega}_q^2)^2 + 4\omega^2\Gamma^2}
\]
where $\mathbf{Q}$ and $\omega$ are the wavevector and energy transfer, respectively, $n(\omega)$ is the Bose function, $\tilde{\omega}_q$ is the phonon energy renormalized by the real part of the phonon self-energy, $\Gamma$ is the intrinsic phonon linewidth closely related to the imaginary part of the phonon self-energy, and $Z(\mathbf{Q})$ is the phonon structure factor. This function covers the energy loss and energy gain scattering by a single line shape. The intensity ratio of the phonon peaks at $E=\pm\omega_q$ is fixed by the principle of detailed balance. Thus, the fit function for the same mode at energy loss and energy gain, e.g., the soft phonon mode [see blue dashed lines in Figs. \ref{fig4}(a)-(e) and \ref{fig5}(a)-(d)], features only three free parameters: $\tilde{\omega}_q$, $\Gamma$ and $Z$. The energy $\omega_q$ of the damped phonons is obtained from the fit parameters of the DHO function by $\omega_q = \sqrt{\tilde{\omega}_q^{\,2} - \Gamma^2}$.

\medskip

The samples were grown using Te self-flux with the addition of small amounts of Pd to the melt as described elsewhere \citep{Straquadine_prb_2019,Ru_prb_2006}. The samples had typical dimensions of $[1 \times 1 \times 0.1]$ mm$^3$. The samples were mounted in a cryostream at 6-ID-D, 30 K $\leq T \leq$ 300 K, and a closed-cycle refrigerator at 30-ID, 10 K $\leq T \leq$ 300 K. The samples were checked against twinning by inspecting the elastic intensities at the wave vectors $\boldsymbol{\mathrm{Q}} = (0, 6, 1)$ and $(1, 6, 0)$, of which only the former is an allowed Bragg reflection. 

We performed \textit{ab-initio} calculations for the lattice dynamical properties based on \textit{density-functional-perturbation-theory} (DFPT) using the high-temperature orthorhombic structure present at $T>T_{1}^{c}$. In RTe$_3$, the $f$ states of the R ion are localized and shifted away from the Fermi energy. Therefore, they are expected to play no direct role in the physics of the $\mathrm{CDW}$ \citep{Brouet_prb_2008,Brouet_prl_2004}. To avoid complications with the $f$ states in standard density-functional theory, we have performed our calculations for LaTe$_3$ in the framework of the mixed basis pseudopotential method \citep{Meyer_fortran_1998}. The exchange-correlation functional was treated in the local-density approximation (LDA). Norm-conserving pseudopotentials were constructed following the scheme of Vanderbilt including 5$s$ and 5$p$ semicore states of La in the valence space but excluding explicitly 4$f$ states \citep{Vanderbilt_prb_1985}. The basis set consisted of plane waves up to 20 Ry complemented with local functions of $s$, $p$, and $d$ symmetry at the La sites. For the exchange-correlation functional the local-density approximation (LDA) was applied \citep{Hedin_jpcssp_1971}. DFPT as implemented in the mixed basis pseudopotential method \citep{Heid_prb_1999} was used to calculate phonon energies and electron-phonon coupling (EPC). An orthorhombic $24\times8\times24$ $\boldsymbol{k}$-point mesh was employed in the phonon calculation, whereas an even denser $48\times12\times48$ mesh was used in the calculation of phonon line widths to ensure proper convergence. This was combined with a standard smearing technique using a Gaussian broadening of 0.2 eV. More details on the calculations can be found elsewhere \citep{Maschek_prb_2018,Maschek_prb_2015}.

\onecolumngrid
\onecolumngrid

\begin{figure}[htb]
\includegraphics[width=\textwidth]{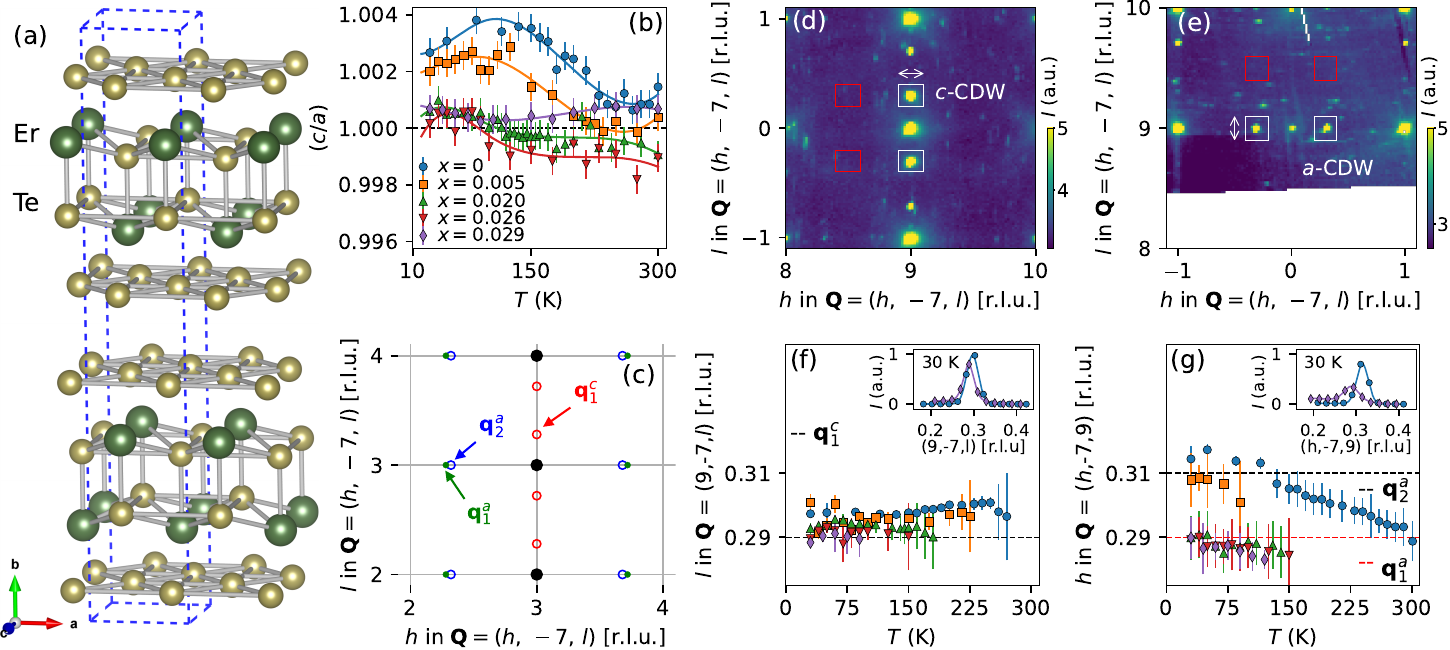}
\caption{\label{fig1} (a) Orthorhombic unit cell of the crystalline structure of $\mathrm{ErTe}_3$ above $T_{1}^{c}$ [Er: large (chartreuse) spheres; Te: small (golden) spheres]. (b) Evolution of the ratio of the $a$ and $c$ in-plane lattice parameters of $\mathrm{Pd}_x\mathrm{ErTe}_3$ as deduced from the XDS data sets. Color-coded solid lines represent the smoothed ratios using spline interpolation. Error bars in (b) are derived from typical uncertainties determining lattice parameters of $2.5 \times 10^{-4}$. (c) Scheme of the reciprocal map indicating positions of the $c\text{-}\mathrm{CDW}$ superlattice peaks at $\boldsymbol{\mathrm{q}}_{1}^{c}$ (red circles) and $a\text{-}\mathrm{CDW}$ superlattice peaks at $\boldsymbol{\mathrm{q}}_{2}^{a}$ (blue circles). Bragg peaks are shown in black dots. Green dots denote the wave vector $\boldsymbol{\mathrm{q}}_{1}^{a}$ for which XDS is observed at high temperature ($x=0$) and in samples with $x \geq 0.02$. (d), (e) Different sections of the reconstructed $(h,-7,l)$ reciprocal plane, indicating the (d) $c\text{-}\mathrm{CDW}$ and (e) $a\text{-}\mathrm{CDW}$ superlattice peaks for pristine $\mathrm{ErTe}_3$ at 30 K. (f), (g) Temperature-dependent incommensurate wave vectors of the (f) $c\text{-}\mathrm{CDW}$ and (g) $a\text{-}\mathrm{CDW}$ (color-coded markers are same as in (b) for different $x$), extracted from the line cuts (see in insets for $x=0,0.29$, color-coded) along $l$ and $h$ directions through the superlattice peaks at $\boldsymbol{\mathrm{Q}}=(9,-7,l)$ and $\boldsymbol{\mathrm{Q}}=(h,-7,9)$ [marked in white boxes in (d), (e), $x=0$], respectively. White arrows in (d), (e) indicate the in-plane binning range for extracting the line cuts from the reciprocal maps. Out-of-plane binning range was $\pm0.5$ r.l.u. along [010]. Horizontal dashed lines in (f), (g) denote $\boldsymbol{\mathrm{q}}_{1}^{c}$, $\boldsymbol{\mathrm{q}}_{2}^{a}$, and $\boldsymbol{\mathrm{q}}_{1}^{a}$ as discussed in the text. White (red) boxes in (d), (e) denote the binning area used to extract temperature dependences of CDW (background) intensities shown in Figs. \ref{fig2}(b)-(f).}
\end{figure}

\twocolumngrid
\twocolumngrid

\section{RESULTS}

\subsection{Diffuse scattering results}

Rare-earth tritellurides feature a nearly perfect square lattice where the in-plane axes $a$ and $c$ differ only by about 0.1\% yielding an orthorhombic structure at room temperature [Fig. \ref{fig1}(a)]. The lattice parameters can be tuned by chemical pressure \citep{Ru_prb_2008_b}. The $c/a$ ratio can also be changed by strain revealing highly responsive CDW properties \cite{Singh_sciadv_2024}. Here, we use x-ray diffraction (XRD) and XDS as powerful tools to study $\mathrm{Pd}_x\mathrm{ErTe}_3$ \citep{Osborn_sciadv_2025} providing access to the dynamics of an incipient CDW transition as well as the properties of the static lattice in the ordered and unorderd state. We find in pristine $\mathrm{ErTe}_3$ that the $c/a$ ratio increases on cooling below $T_{1}^{c}$ and decreases on further cooling to below $T_{2}^{a}$ [Fig. \ref{fig1}(b)]. Similarly, we can identify the corresponding transition temperatures in the $c/a$ ratio for Pd$_{0.005}$ErTe$_3$. For larger intercalation levels $x \geq 0.02$, we observe much weaker but clearly visible upturns in the $c/a$ below the reported $T_{1}^{c}$ transition temperatures \citep{Straquadine_prb_2019} except for our maximum intercalation level, $x = 0.029$, for which a previous analysis of our XDS data already suggested that $T_{1}^{c}$ is suppressed to zero \citep{Mallayya_nphys_2024}.

Figure \ref{fig1}(c) shows a plane in the reciprocal space indicating the superlattice peak positions of the $c\text{-}\mathrm{CDW}$ ($\boldsymbol{\mathrm{q}}_{1}^{c}$, red circles) and the $a\text{-}\mathrm{CDW}$ ($\boldsymbol{\mathrm{q}}_{2}^{a}$, blue circles) with respect to fundamental Bragg peaks of the orthorhombic high-temperature structure (black dots). As shown below, XDS along $a^*$ appears at a wave vector $\boldsymbol{\mathrm{q}}_{1}^{a}$ (green dots) which is slightly different from that of the $a\text{-}\mathrm{CDW}$, $\boldsymbol{\mathrm{q}}_{2}^{a}$. Correspondingly, CDW superlattice peaks in pristine $\mathrm{ErTe}_3$ are observed at $\boldsymbol{\mathrm{q}}_{1}^{c}=(0,0,0.3)$ [marked by the white boxes in Fig. \ref{fig1}(d)] and $\boldsymbol{\mathrm{q}}_{2}^{a}=(0.31,0,0)$ [marked by the white boxes in Fig. \ref{fig1}(e)] below the transition temperatures $T_{1}^{c}=270$ K and $T_{2}^{a}=160$ K, respectively. In the following we are interested in the detailed temperature evolution of the wavevector values for which we find XDS above and static superlattice peak below the respective CDW ordering temperatures. To this end we analyze line scans along the [001] ($c\text{-}\mathrm{CDW}$) and [100] ($a\text{-}\mathrm{CDW}$) directions in reciprocal space extracted from the 3D XDS data sets with binning in the orthogonal directions of $\pm0.1$ r.l.u. within [see white arrows in Figs. \ref{fig1}(d), (e)] and $\pm0.5$ r.l.u. perpendicular to the basal plane. Positions of at least two fundamental Bragg peaks in one scan are fixed to integer positions and, relative to that, we define the incommensurate positions of the CDW ordering wave vectors. Exemplary XDS data are shown in the insets of Figs. \ref{fig1}(f) (determining $\boldsymbol{\mathrm{q}}_{1}^{c}$) and (g) (determining $\boldsymbol{\mathrm{q}}_{1}^{a}$/$\boldsymbol{\mathrm{q}}_{2}^{a}$). The respective main panels show the deduced temperature dependences of superlattice peaks at $\boldsymbol{\mathrm{q}} = (0,0,l)$ and $(h,0,0)$ in Figs. \ref{fig1}(f) and (g), respectively. We report results for all temperatures which provided reasonable error bars for XDS above or static scattering below the transition temperatures. Unfortunately, the data quality for $x = 0.005$ did not allow to extract reasonable results at temperatures significantly above the CDW transition temperatures. In the following, we focus on the analysis of the wave vector position of XDS above and CDW superlattice peaks below the respective transition temperatures. Details of the evolution of the linewidths of peaks in XDS based on the same data used for our analysis were discussed in a previous publication \citep{Mallayya_nphys_2024}. Here, we point out that the XDS data were taken in a high-throughput measurement using high x-ray energies of 87 keV.  These measurements have lower resolution than is typically used for linewidth studies but allowed using a novel peak-spread analysis to assess the linewidths of the diffuse CDW peaks from of a large number of peaks ($\sim 3000$), providing selection-bias free results and allowing to obtain the intrinsic linewidth from the momentum dependence \citep{Mallayya_nphys_2024}.

We find different evolutions of the XDS and superlattice-peak wave vectors along the two in-plane directions. For $\boldsymbol{\mathrm{q}} = (0,0,l)$, the peak is observed at $l=0.3-0.29$ and shows a gradual but small evolution for increasing intercalation values [Fig. \ref{fig1}(f)]. For $\boldsymbol{\mathrm{q}} = (h,0,0)$, we find $h=0.31$ for $x = 0, 0.005$ but $h=0.29$ for all samples with $x \geq 0.02$ [Fig. \ref{fig1}(g)]. This change of the wave vector position from samples with $x \leq 0.01$ to samples with $x \geq 0.02$ (see insets in Figs. \ref{fig1}(f)(g)) is confirmed by high-resolution momentum scans at zero energy transfer performed at the HERIX spectrometer [see green symbols in the insets of Figs. \ref{fig4}(f), $x = 0.01$, and \ref{fig5}(e), $x = 0.023$]. Interestingly, XDS at $\boldsymbol{\mathrm{q}} = (h,0,0)$ in pristine $\mathrm{ErTe}_3$ is observed at $h = 0.29$ at room temperature and only gradually moves to $h = 0.31$ on cooling [blue dots in Fig. \ref{fig1}(g)], where we find $a\text{-}\mathrm{CDW}$ superlattice peaks for $T \leq T_2^a=160$ K. Here, we introduce the new wave vector definition $\boldsymbol{\mathrm{q}}_{1}^{a}=(0.29,0,0)$ to distinguish the wave vector position of XDS in highly intercalated samples and pristine $\mathrm{ErTe}_3$ at room temperature from that of superlattice peaks at $\boldsymbol{\mathrm{q}}_{2}^{a}=(0.31,0,0)$ indicating the presence of $a\text{-}\mathrm{CDW}$ order [see map of momentum space in panel Fig. \ref{fig1}(c)]. We note that all peaks centered at $\boldsymbol{\mathrm{q}}_{1}^{a}$ are not resolution limited and, thus, reflect only short-to-medium-range correlations along the $a$ axis in $\mathrm{Pd}_x\mathrm{ErTe}_3$. We argue below that the $a\text{-}\mathrm{CDW}$ transition with its characteristic wave vector position $\boldsymbol{\mathrm{q}}_{2}^{a}$ is indeed suppressed at intercalation levels of $x\geq0.02$ and that the residual XDS at $\boldsymbol{\mathrm{q}}_{1}^{a}$ originates from a partial phonon softening directly related to the onset of the $c\text{-}\mathrm{CDW}$ at $\boldsymbol{\mathrm{q}}_{1}^{c}$.

\begin{figure}[!t]
\includegraphics[width=0.475\textwidth]{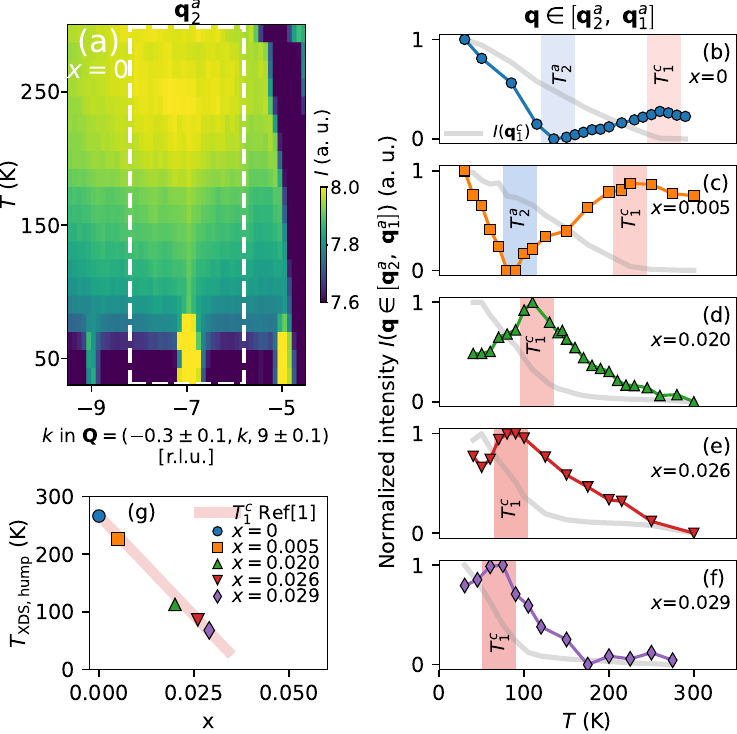}
\caption{\label{fig2} (a) Integrated XDS intensity along $\boldsymbol{\mathrm{Q}} = (-0.3(\pm0.1), k, 9(\pm0.1))$, $T = 30$ – 300 K obtained in $\mathrm{ErTe}_3$, reveal a considerable increase near $\boldsymbol{\mathrm{q}}_{2}^{a}$ around 260 K, i.e., close to $T_{1}^{c}$. (b)-(f) Integrated intensities of XDS (symbols) around $\boldsymbol{\mathrm{Q}} = (-0.3(\pm 0.1), -7(\pm0.5), 9(\pm0.1))$ (includes $\boldsymbol{\mathrm{q}}_{2}^{a}$ and $\boldsymbol{\mathrm{q}}_{1}^{a}$) for $x = 0$ (b), 0.005 (c), 0.020 (d), 0.026 (e), and 0.029 (f) and temperatures $T = 30$ - 300 K. Background scattering [$\boldsymbol{\mathrm{Q}}$ area indicated by red boxes in Fig. \ref{fig1}(e)] was subtracted. The XDS integrated intensities [$\boldsymbol{\mathrm{Q}}$ area indicated by white boxes in Fig. \ref{fig1}(e)] are normalized to a $[0-1]$ scale. Grey-shaded lines indicate corresponding results [$\boldsymbol{\mathrm{Q}}$ areas indicated by white and red boxes in Fig. \ref{fig1}(d)] at the orthogonal wave vectors $\boldsymbol{\mathrm{Q}} = (9(\pm0.1), -0.3(\pm 0.1), -7(\pm0.5))$ (including $\boldsymbol{\mathrm{q}}_{1}^{c}$). Red and blue vertical shades mark maxima of XDS intensity which coincide with the respective doping dependent values of $T_{1}^{c}$ and $T_{2}^{a}$, respectively [see \citep{Straquadine_prb_2019}]). (g) Color-coded symbols represent the temperatures corresponding to the hump in the XDS intensities in Figs. (b-f). Red shaded line represents a linear guide to $T_{1}^{c}$ as reported in Ref. \citep{Straquadine_prb_2019}.}
\end{figure}

Now we look at the temperature dependence of XDS along the [100] direction [Fig. \ref{fig2}(a)]. Integrating over a volume in reciprocal space including $\boldsymbol{\mathrm{q}}_{2}^{a}$ and $\boldsymbol{\mathrm{q}}_{1}^{a}$ $([\Delta h\times\Delta k\times\Delta l]=[0.25\times0.25\times0.25] \mathrm{\AA}^{-3})$, we clearly see the onset of the secondary CDW transition in low-intercalated samples at $T_2^a$ [blue shaded vertical bars in Figs. \ref{fig2}(b) and \ref{fig2}(c)]. Yet, XDS extends all the way up to room temperature. Interestingly, the detailed temperature dependence of the intensity around $\boldsymbol{\mathrm{Q}}=(-0.3,-7,9)$ [white box in Fig. \ref{fig2}(a)] reveals a broad hump of XDS intensity centered at $T_{1}^{c}=260$ K for $x = 0$ [red-shaded vertical bars in Fig. \ref{fig2}(b)] and $T_{1}^{c}=230$ K for $x = 0.005$ [red shaded vertical bars in Fig. \ref{fig2}(c)]. Here, the values of $T_{1}^{c}$ are determined by the analysis of similarly integrated scattering around $\boldsymbol{\mathrm{q}}_{1}^{c}$ (grey-shaded lines) and the obtained values are in good agreement with previous reports \citep{Straquadine_prb_2019}. At higher intercalation levels, this broad peak still evolves according to the reported values of $T_{1}^{c}$) as function of Pd-intercalation [Figs. \ref{fig2}(d)-(f)]. On the other hand, a clear signature of the $a\text{-}\mathrm{CDW}$ transition is missing for samples with $x\geq0.02$. XDS integrates over the typical energy range of lattice vibrations in solids where low-energy phonons contribute most strongly to the XDS because the one-phonon scattering cross section scales with the inverse of the phonon energy and because of the thermal phonon occupation which also leads to increasing intensities when the phonon energy is reduced \citep{baron_arxiv_2020}. Strongly temperature-dependent XDS is therefore often associated with phonon softening and, indeed, XDS, is routinely used to identify spots in momentum space at which more detailed phonon studies are most promising \citep{Souliou_prl_2022,LeTacon_nphys_2014}. In $\mathrm{Pd}_x\mathrm{ErTe}_3$, the apparent link of the maximum of XDS with the intercalation dependence of $T_1^c$ [Fig. \ref{fig2}(g)] suggests that phonon softening occurs along both in-plane directions simultaneously.

\subsection{Phonons from IXS}  
\begin{figure}[!t] 
\includegraphics[width=0.475\textwidth]{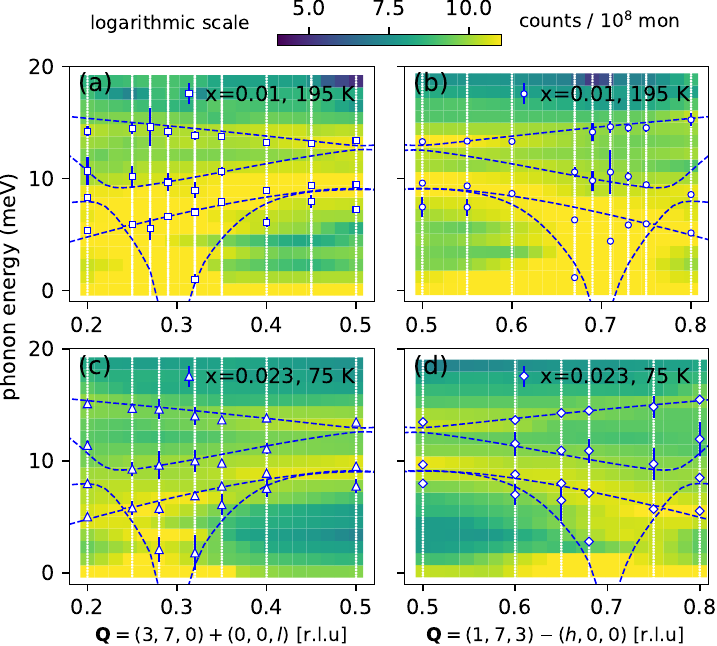}
\caption{\label{fig3} Comparison of measured (symbols) and calculated phonon dispersions (dashed lines) in $\mathrm{Pd}_x\mathrm{ErTe}_3$ with (a),(b) $x=0.01$, $T = 195$ K, and (c),(d) $x=0.023$, $T = 75$ K across $\boldsymbol{\mathrm{q}}_{1}^{c}$ (left panels) and $\boldsymbol{\mathrm{q}}_{2}^{a}$/$\boldsymbol{\mathrm{q}}_{1}^{a}$ (right panels). Raw IXS intensities are shown in color-code (log scale). Vertical lines denote $\boldsymbol{\mathrm{Q}}$ values of performed IXS scans. Symbols represent approximated phonons energies where exemplary fits are shown in Figs. \ref{fig4} and \ref{fig5}.}
\end{figure}

Here, we investigated the lattice dynamical properties of Pd-intercalated $\mathrm{ErTe}_3$ with meV-resolution IXS \citep{Said_jsr_2020} on two samples with $x=0.01$ and $0.023$. In analogy to previous studies of TbTe$_3$ \citep{Maschek_prb_2015} and DyTe$_3$ \citep{Maschek_prb_2018}, we measured the phonon dispersions across both $\boldsymbol{\mathrm{q}}_{1}^{c}$ and $\boldsymbol{\mathrm{q}}_{2}^{a}$/$\boldsymbol{\mathrm{q}}_{1}^{a}$ for both samples in the Brillouin zones adjacent to the reciprocal lattice vectors $\boldsymbol{\tau} = (3,7,0)$ and $(1,7,3)$ [\textcolor{blue}{see note} \footnote{Note that $\boldsymbol{\mathrm{Q}} = (0,7,3)$ is not an allowed reflection but corresponds to a zone boundary in the Brillouin zone of $\mathrm{Pd}_x\mathrm{ErTe}_3$}]. We note that the momentum resolution of our standard phonon scans with the circular opening of the analyzer d$_\mathrm{ana}=95$ mm was not sufficient to resolve the difference between $\boldsymbol{\mathrm{q}}_{2}^{a}$ and $\boldsymbol{\mathrm{q}}_{1}^{a}$. Scattering data taken near the corresponding values of $T_{1}^{c}$ for $x = 0.01$ [Figs. \ref{fig3}(a) and \ref{fig3}(b)] and 0.023 [Figs. \ref{fig3}(c) and \ref{fig3}(d)] reveal several phonon branches propagating along the [001] and [100] directions (symbols in Fig. \ref{fig3}) and the deduced phonon energies are in reasonable agreement with \textit{ab-initio} lattice dynamical calculations based on density functional perturbation theory (dashed lines in Fig. \ref{fig3}). The calculations are the same as presented in our work on DyTe$_3$ except for a scaling taking into account the different masses of Er and Dy. Here, we only briefly summarize the main features of the calculations and refer the interested reader to Ref. \citep{Maschek_prb_2018} for a more detailed description and discussion. We present $T = 0$ K, i.e., ground state calculations which correctly reflect the CDW structural instabilities at the CDW ordering wave vectors along the $c^*$ [Fig. \ref{fig3}(a) and (c)] and $a^*$ directions [Fig. \ref{fig3}(b) and (d)]. Here, the calculated imaginary phonon energies are displayed as negative values. The leading soft mode, i.e., instability is predicted to be along the $c^*$ direction in agreement with experiment. Intercalation cannot be taken into account in our calculations since very large supercells would be required to describe intercalation levels of few percent. On the other hand, we do not expect such small concentrations to significantly alter the lattice dynamical properties except for the soft phonon modes.

\begin{figure}[!htb] 
\includegraphics[width=0.475\textwidth]{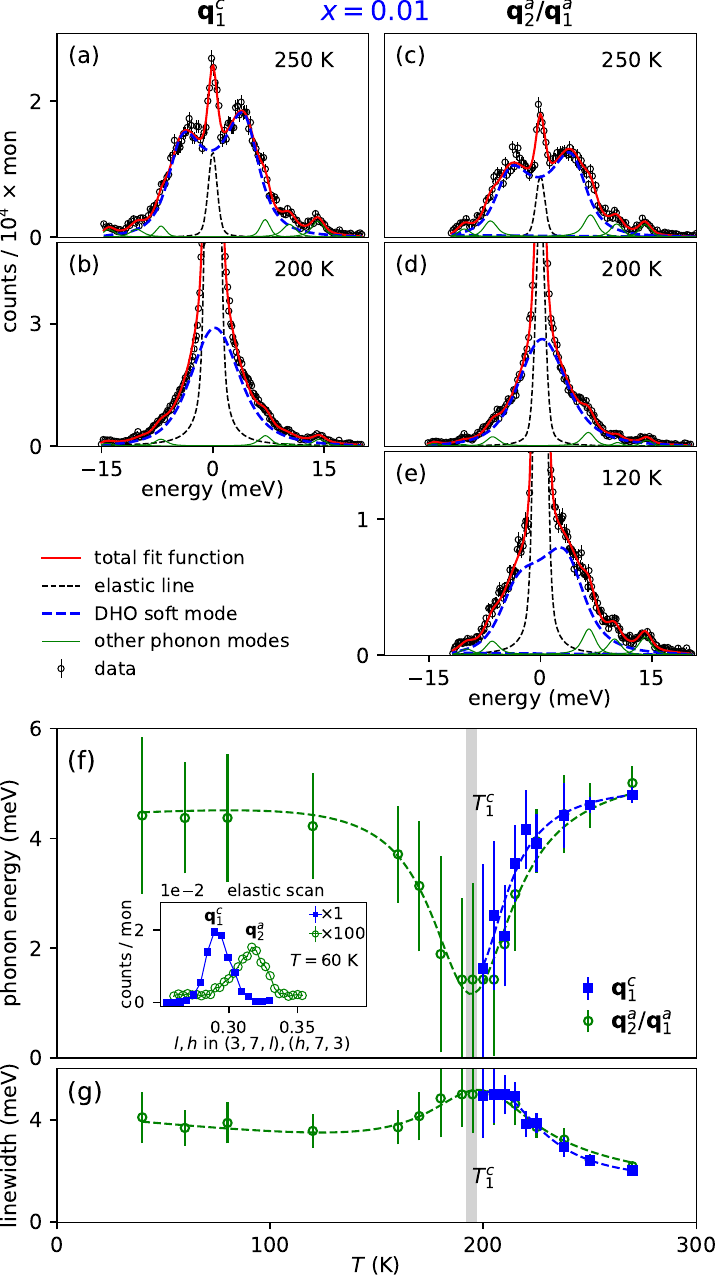}
\caption{\label{fig4} Phonon spectroscopy in $\mathrm{Pd}_{0.01}\mathrm{ErTe}_3$. Constant-$\boldsymbol{\mathrm{Q}}$ scans (symbols) at wave vectors (a), (b) $\boldsymbol{\mathrm{Q}} = (3,7,0.29)$ and (c)-(e) $\boldsymbol{\mathrm{Q}} = (0.31,7,3)$ corresponding to $\boldsymbol{\mathrm{q}}_{1}^{c}$ and $\boldsymbol{\mathrm{q}}_{2}^{a}$/$\boldsymbol{\mathrm{q}}_{1}^{a}$ (see text), respectively. Solid lines in red represent fits to the data consisting of damped harmonic oscillator (DHO) functions convoluted with the experimental resolution for the soft phonon mode (blue dashed) and other phonon modes (green solid), a sloping background (too small to be visible) and a resolution-limited pseudo-Voigt function for the elastic line (black dashed). (f), (g) Derived temperature dependence of the respective energies and linewidths of the soft phonon mode at $\boldsymbol{\mathrm{q}}_{1}^{c}$ and $\boldsymbol{\mathrm{q}}_{2}^{a}$/$\boldsymbol{\mathrm{q}}_{1}^{a}$ for $\mathrm{Pd}_{0.01}\mathrm{ErTe}_3$. The inset in (f) shows high-resolution momentum scans at zero energy transfer across $\boldsymbol{\mathrm{q}}_{1}^{c}$ and $\boldsymbol{\mathrm{q}}_{2}^{a}$/$\boldsymbol{\mathrm{q}}_{1}^{a}$ at $T=60$ K. The vertical grey-shaded bar indicates the transition temperature $T_{1}^{c} = 195$ K reported for $\mathrm{Pd}_{0.01}\mathrm{ErTe}_3$ \citep{Straquadine_prb_2019}.}
\end{figure}

\begin{figure}[!htb] 
\includegraphics[width=0.475\textwidth]{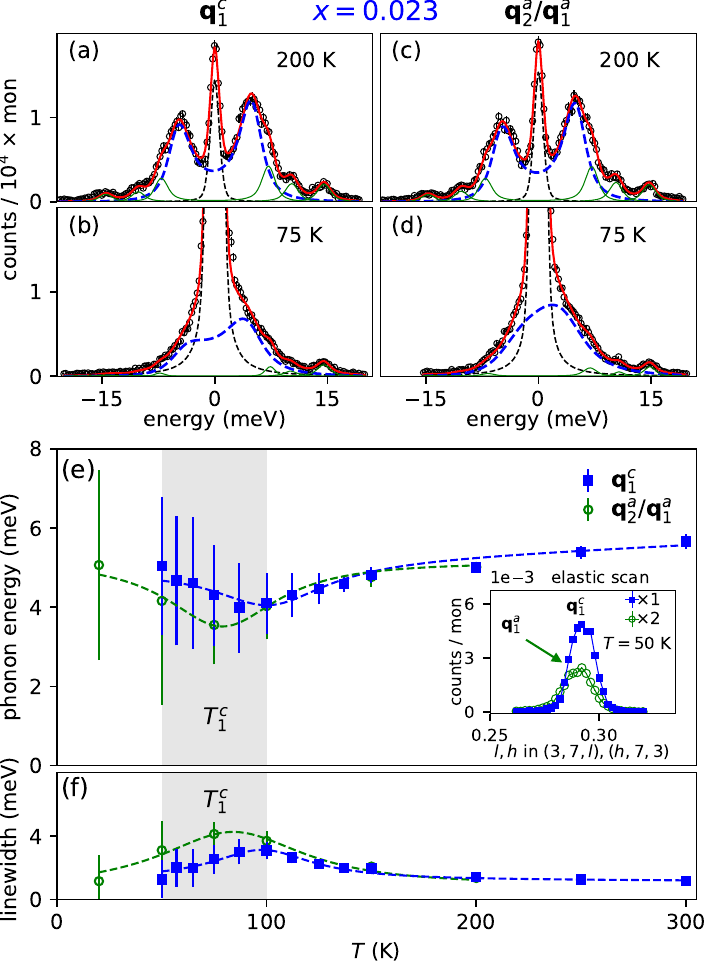}
\caption{\label{fig5}Phonon spectroscopy in $\mathrm{Pd}_{0.023}\mathrm{ErTe}_3$. Constant-$\boldsymbol{\mathrm{Q}}$ scans (symbols) at wave vectors (a), (b) $\boldsymbol{\mathrm{Q}} = (3,7,0.29)$ and (c), (d) $\boldsymbol{\mathrm{Q}} = (0.29,7,3)$ corresponding to $\boldsymbol{\mathrm{q}}_{1}^{c}$ and $\boldsymbol{\mathrm{q}}_{2}^{a}$/$\boldsymbol{\mathrm{q}}_{1}^{a}$ (see text), respectively. Symbol/line code is the same as in Fig. \ref{fig4}. (e), (f) Derived temperature dependence of the respective energies and linewidths of the soft phonon modes at $\boldsymbol{\mathrm{q}}_{1}^{c}$ and $\boldsymbol{\mathrm{q}}_{2}^{a}$/$\boldsymbol{\mathrm{q}}_{1}^{a}$ for $\mathrm{Pd}_{0.023}\mathrm{ErTe}_3$. The inset in (e) shows high-resolution momentum scans at zero energy transfer across $\boldsymbol{\mathrm{q}}_{1}^{c}$ and $\boldsymbol{\mathrm{q}}_{2}^{a}$/$\boldsymbol{\mathrm{q}}_{1}^{a}$ at $T=50$ K. The vertical grey shaded bar indicates the temperature range reported for $T_{1}^{c} = 50$ - 100 K in $\mathrm{Pd}_{0.023}\mathrm{ErTe}_3$ \citep{Straquadine_prb_2019}.}
\end{figure}

In the following we focus on the temperature dependence of the soft modes measured at $\boldsymbol{\mathrm{q}}_{1}^{c}$ and $\boldsymbol{\mathrm{q}}_{2}^{a}$/$\boldsymbol{\mathrm{q}}_{1}^{a}$. A typical energy scan taken well above the transition temperature $T_{1}^{c} = 195$ K for an intercalation level of $x = 0.01$ shows the CDW soft mode at an energy of about 5 meV [blue dashed line in Fig. \ref{fig4}(a)] along with three other weaker phonon peaks (green solid lines). Similar to previous phonon measurements in $\mathrm{TbTe}_3$ \citep{Maschek_prb_2015} and $\mathrm{DyTe}_3$ \citep{Maschek_prb_2018}, we found that only the soft mode shows a strong temperature dependence. At $T = 200$ K, just above $T_{1}^{c}$, the soft mode at $\boldsymbol{\mathrm{q}}_{1}^{c}$ softens close to zero energy [Fig. \ref{fig4}(b)]. The intensities of both the soft phonon mode and the elastic contribution are much larger than those observed at $T = 250$ K [see Fig. \ref{fig4}(a)]. Interestingly, we have near-identical behaviour of the lowest-energy mode at $\boldsymbol{\mathrm{q}}_{2}^{a}$/$\boldsymbol{\mathrm{q}}_{1}^{a}$ [Figs. \ref{fig4}(c) and \ref{fig4}(d)]: The most intense mode softens from 5 meV ($T = 250$ K) to close to zero energy ($T = 200$ K) accompanied by a strong increase of the phonon and elastic line intensities. However, we can follow the evolution of the quasi-soft mode at $\boldsymbol{\mathrm{q}}_{2}^{a}$/$\boldsymbol{\mathrm{q}}_{1}^{a}$ on cooling below $T_{1}^{c}$, since no CDW superstructure peak is developing [\textcolor{blue}{see note} \footnote{The strong Lorentzian character of the energy resolution function in IXS yields an overpowering high background for all phonon energies at any wave vector featuring a Bragg peak. This includes superlattice peaks of a long-range ordered CDW although these are about three orders of magnitude less intense than fundamental Bragg peaks in $\mathrm{RTe}_3$ \citep{Ru_prb_2008_b}}]. Data taken at $T = 120$ K demonstrate a clear hardening of the quasi-soft mode at $\boldsymbol{\mathrm{q}}_{2}^{a}$/$\boldsymbol{\mathrm{q}}_{1}^{a}$ [Fig. \ref{fig4}(e)]. Both soft modes, at $\boldsymbol{\mathrm{q}}_{1}^{c}$ and $\boldsymbol{\mathrm{q}}_{2}^{a}$/$\boldsymbol{\mathrm{q}}_{1}^{a}$, show within the error bar identical temperature dependences of the phonon energy above $T_{1}^{c}$ [Fig. \ref{fig4}(f)]. We do not see any softening of the phonon at $\boldsymbol{\mathrm{q}}_{2}^{a}$/$\boldsymbol{\mathrm{q}}_{1}^{a}$ on cooling further down to 50 K [green symbols in Fig. \ref{fig4}(f)] which is close to the expected value of $T_{2}^{a}$ for $x = 0.01$ \citep{Straquadine_prb_2019}. Further, we performed momentum scans at zero energy transfer with the circular opening of the analyzer reduced from $\mathrm{d}_\mathrm{ana}$ = 95 mm to 15 mm to resolve the difference between $\boldsymbol{\mathrm{q}}_{2}^{a}$ and $\boldsymbol{\mathrm{q}}_{1}^{a}$ and found the scattering to be centered on $\boldsymbol{\mathrm{q}}_{2}^{a}$ [see inset in Fig. \ref{fig4}(f)] showing that Pd$_{0.01}$ErTe$_{3}$ still features strong $a\text{-}\mathrm{CDW}$ correlations similar to our observation for $x = 0.005$ in XDS data [see Fig. \ref{fig1}(g)]. 

Our IXS measurements explain that XDS at $\boldsymbol{\mathrm{q}}_{2}^{a}$/$\boldsymbol{\mathrm{q}}_{1}^{a}$ shows a maximum in its intensity at $T_1^c$ because the phonon contribution to XDS scales with the inverse of its energy. The observed temperature dependent phonon softening at $\boldsymbol{\mathrm{q}}_{2}^{a}$/$\boldsymbol{\mathrm{q}}_{1}^{a}$ [$E_\mathrm{phonon}$($T = 270$ K) = 5 meV to $E_\mathrm{phonon}$($T = 200$ K) = 1.4 meV; see Fig. \ref{fig4}(f)] results in an increase of its intensity by a factor of 9 taking into account the reduced phonon energy (increase by a factor 3.6) as well as the temperature and the energy dependence of the Bose factor (increase by factor 2.5) on cooling.

We chose $x=0.023$ as second intercalation level to be investigated by IXS. For this intercalation level, $T_{1}^{c}$ is suppressed to 50 - 100 K and superconductivity sets in below $T_\mathrm{sc}=2.6$ K \citep{Straquadine_prb_2019}. Indeed, a set of elastic scans taken during the presented IXS study (done with $\mathrm{d}_\mathrm{ana}=15$ mm) and its analysis (\textcolor{blue}{already published \citep{Straquadine_prb_2019}}) demonstrate that CDW long-range order is observed at $\boldsymbol{\mathrm{q}}_{1}^{c}$ whereas elastic scattering at $\boldsymbol{\mathrm{q}}_{1}^{a}$ does not become resolution limited down to 20 K (see Fig. 7 in Ref. \citep{Straquadine_prb_2019}). It was already noted in \citep{Straquadine_prb_2019} that the elastic scattering along [100] appears at $\boldsymbol{\mathrm{q}}_{1}^{a} =(0.29,0,0)$, and not $\boldsymbol{\mathrm{q}}_{2}^{a}=(0.31,0,0)$[see inset of Fig. \ref{fig5}(e)]. This is in agreement with our XDS results for similar doping levels [see Fig. \ref{fig1}(g)]. 

Guided by the results from the elastic high-resolution momentum scans, we performed detailed studies of the phonon softening at $\boldsymbol{\mathrm{q}}_{1}^{c}=(0,0,0.29)$ and $\boldsymbol{\mathrm{q}}_{1}^{a}=(0.29,0,0)$. Phonon scan at $T=200$ K and 75 K ($\approx T_1^c$) reveal phonon softening [Figs. \ref{fig5}(a)-(d)]. However, soft phonons at both wave vectors remain always at a finite energy with minima of about 3-4 meV located below 100 K [Fig. \ref{fig5}(e)]. The deduced phonon energies indicate a hardening of the soft mode energies for the lowest temperatures at both wave vectors. However, error bars increase significantly due to increased elastic intensities.

In comparison, there are similarities but also clear differences in the way phonons soften in $\mathrm{Pd}_x\mathrm{ErTe}_3$ for $x=0.01$ [Fig. \ref{fig4}(f)] and 0.023 [Fig. \ref{fig5}(e)]. A similarity is that we observe softening at the respective values of $T_1^c$ in both samples at $\boldsymbol{\mathrm{q}}_{1}^{c}$ and $\boldsymbol{\mathrm{q}}_{2}^{a}$/$\boldsymbol{\mathrm{q}}_{1}^{a}$. The softening observed at the two different wave vectors for each intercalation level is within the error bar the same although the softening is much weaker for $x=0.023$.

\section{DISCUSSION}
Taken together, our analysis provides compelling evidence that the XDS observed near $\boldsymbol{\mathrm{q}}_{1}^{a}$ and $T_{1}^{c}$ originates in a partial phonon softening and is reflected in our calculations where two soft phonon modes compete (see Fig. \ref{fig3} and extended discussion in \citep{Maschek_prb_2018}. The leading soft mode is located near $\boldsymbol{\mathrm{q}}_{1}^{c}$ in agreement with experiment. In contrast to phonon measurements in pristine $\mathrm{DyTe}_3$ \citep{Maschek_prb_2018}, the degree of softening in intercalated samples at $\boldsymbol{\mathrm{q}}_{1}^{c}$ and $\boldsymbol{\mathrm{q}}_{2}^{a}$ cannot be distinguished anymore. At the same time, the softening is clearly incomplete for $x=0.023$. The observed shift in the XDS from $\boldsymbol{\mathrm{q}}_{2}^{a}$, $x \leq 0.005$, to $\boldsymbol{\mathrm{q}}_{1}^{a}$, $x \geq 0.02$ [Fig. \ref{fig1}(f)], the evolution of the bump in the XDS with intercalation [Figs. \ref{fig2}(b)-(f)] and the reported phonon properties suggest that previously reported remnants of the $a\text{-}\mathrm{CDW}$ in highly intercalated $\mathrm{Pd}_x\mathrm{ErTe}_3$ \citep{Fang_prb_2019} are rather signatures of the competing soft phonon modes associated with the $c\text{-}\mathrm{CDW}$.   

We argue that the momentum position of XDS at room temperature in pristine $\mathrm{ErTe}_3$ at $\boldsymbol{\mathrm{q}}_{1}^{a}$ [see Fig. \ref{fig1}(g)] is reasonable in that the undistorted structure along with its Fermi surface is nearly tetragonal and, thus, the competing soft phonon modes should occur at nearly identical wave vectors. Our own previous results from phonon spectroscopy in DyTe$_3$ \citep{Maschek_prb_2018} simply did not have good enough momentum resolution and were guided by the momentum position of the $a\text{-}\mathrm{CDW}$ at low temperatures.

Our study is in line with several recent reports on Pd-intercalated $\mathrm{ErTe}_3$ \citep{Mallayya_nphys_2024,Straquadine_prb_2019,Fang_prb_2019,Singh_sciadv_2024,Corasaniti_prr_2023}. STM measurements \citep{Fang_prb_2019} showed that intercalation induces CDW dislocations which have a stronger effect on the secondary CDW order along the $a^*$ direction explaining the quick suppression of this order at low intercalation levels \citep{Straquadine_prb_2019}. Surprisingly, measurements at high intercalation levels, $x = 0.05$, seem to show vestiges of both distinct CDW phases, far beyond where signatures of the smeared phase transitions are observed in bulk probes. Our analysis suggests that the $a\text{-}\mathrm{CDW}$ order is completely suppressed near $x = 0.01$ and that scattering in highly intercalated samples near $\boldsymbol{\mathrm{q}}_{2}^{a}$ is, in fact, related to the persisting strong competition of the soft phonon modes along both in-plane directions. This is also consistent with the arguments that intercalation and the induced CDW dislocations help restoring the \textit{normal-state} of the Fermi surface \citep{Kogar_nphys_2020,Fang_prb_2019} which in turn is subject to strong competition of the soft phonon modes as we observe it close to $T_{1}^{c}$. Thus, the proposed phase diagram for high intercalation levels \citep{Fang_prb_2019} has to be revised in that our results for $x \leq 0.029$ suggest that no remnants of the $a\text{-}\mathrm{CDW}$ order survive. That intercalated samples are mostly dominated by features related to the $c\text{-}\mathrm{CDW}$ transition seems to be consistent with the observation that the nematic elastoresistance of Pd-intercalated $\mathrm{ErTe}_3$ features a more symmetric response than observed in pristine $\mathrm{ErTe}_3$ \citep{Mallayya_nphys_2024,Singh_sciadv_2024}. Compared to previous measurements in pristine $\mathrm{DyTe}_3$ \citep{Maschek_prb_2018}, the phonon softening observed along $c^*$ and $a^*$ approach each other regarding its temperature dependence. The softening is clearly incomplete for $x = 0.023$. It is an interesting question for future studies whether the incomplete softening is simply an effect of disorder or related to the evolution of the $c\text{-}\mathrm{CDW}$ phase in pristine $\mathrm{ErTe}_3$ to a CDW Bragg glass in intercalated $\mathrm{Pd}_x\mathrm{ErTe}_3$ as reported recently \citep{Mallayya_nphys_2024}. The incomplete phonon softening is in contrast to observations in $2H\text{-}\mathrm{TaSe}_2$ when the CDW order is suppressed by pressure. In this case without disorder, a (within the error bar) complete phonon softening is observed when $T_\mathrm{CDW}$ is suppressed to zero near a pressure of 20 GPa \citep{Tymoshenko_comphys_2025}. 

It is also instructive to compare our results with a non-equilibrium study of $\mathrm{LaTe}_3$ \citep{Kogar_nphys_2020}. In equilibrium, $\mathrm{LaTe}_3$ features only the $c\text{-}\mathrm{CDW}$ at $\boldsymbol{\mathrm{q}}_{1}^{c} = (0,0,0.29)$ \citep{Ru_prb_2008_b}. After photoexcitation, superlattice peaks are observed in electron diffraction at $\boldsymbol{\mathrm{q}}_{1}^{a} = (0.29,0,0)$. The opposite time dependent diffraction intensities at $\boldsymbol{\mathrm{q}}_{1}^{c}$ and $\boldsymbol{\mathrm{q}}_{1}^{a}$ are evidenced for a strong competition and were, indeed, discussed with regard to the earlier reported soft mode competition in $\mathrm{DyTe}_3$ \citep{Maschek_prb_2018}. $\mathrm{Pd}_x\mathrm{ErTe}_3$ with $x > 0.01$ can be seen as an equilibrium analogue to photoexcited $\mathrm{LaTe}_3$: Similar to the latter, our XDS data for $x = 0.02$ and the elastic scans for $x=0.023$ [see inset in Fig. \ref{fig5}(e)] reveal the presence of a significant scattering intensities along both in-plane orthogonal directions at $\boldsymbol{\mathrm{q}}_{1}^{c}$ and $\boldsymbol{\mathrm{q}}_{1}^{a}$. The difference of the corresponding x-ray scattering intensities, deduced from the elastic scans, is fairly small (factor of 5 for $x = 0.023$, see Fig. 7 in \citep{Maschek_prb_2018}).

Finally, our results for $x = 0.01$ are consistent with our previous measurements in $\mathrm{DyTe}_{3}$ \citep{Maschek_prb_2018} in that we do not observe any phonon softening below $T_{1}^{c}$ at $\boldsymbol{\mathrm{q}}_{2}^{a}$/$\boldsymbol{\mathrm{q}}_{1}^{a}$ [see Fig. \ref{fig4}(f)]. The absence of phonon softening is evidence that $a\text{-}\mathrm{CDW}$ order is not simply a $90^{\circ}$-rotated analogue of the $c\text{-}\mathrm{CDW}$. This conclusion was also put forward by a recent time-resolved ARPES study reporting qualitatively different responses of the two CDW orders in $\mathrm{ErTe}_3$ to light excitation \citep{Su_arXiv_2025}. Yet, a recent Raman scattering study on $\mathrm{ErTe}_3$ and $\mathrm{HoTe}_3$ \citep{SinghB_nphys_2025} observed amplitude mode softening on heating towards $T_{2}^{a}$ within the $a\text{-}\mathrm{CDW}$ phase, implicitly contradicting our current and previous \citep{Maschek_prb_2018} x-ray scattering and other optical measurements \citep{Lavagnini_prb_2010, Lavagnini_prb_2008, Yumigeta_aplmaterials_2022,Pfuner_prb_2010}.
Thus, the CDW formation mechanism for the $a\text{-}\mathrm{CDW}$ order in $\mathrm{ErTe}_3$ remains debated though some evidence points towards a mechanism distinct from that of the $c\text{-}\mathrm{CDW}$.

\section{CONCLUSION}

In summary, our study highlights the evolution of the CDW soft phonons in $\mathrm{Pd}_x\mathrm{ErTe}_3$ by means of XDS and IXS. We show that the transition in pristine $\mathrm{ErTe}_3$ into the primary $c\text{-}\mathrm{CDW}$ phase, with $\boldsymbol{\mathrm{q}}_{1}^{c}$ along $c^*$, is accompanied by increased XDS at the ordering wave vector of the secondary $a\text{-}\mathrm{CDW}$ phase, $\boldsymbol{\mathrm{q}}_{2}^{a}$ along $a^*$. At higher intercalation levels, this XDS along $a^*$ shifts to $\boldsymbol{\mathrm{q}}_{1}^{a}$, where the absolute value of $\boldsymbol{\mathrm{q}}_{1}^{a}$ and $\boldsymbol{\mathrm{q}}_{1}^{c}$ are equivalent, within the error bar of the experiment. Phonon spectroscopy using IXS reveals the competing soft phonon modes at $\boldsymbol{\mathrm{q}}_{1}^{c}$ on the one side and $\boldsymbol{\mathrm{q}}_{2}^{a}$ or $\boldsymbol{\mathrm{q}}_{1}^{a}$, respectively, on the other side. The phonon energies do not soften to zero energy at $x = 0.023$ in contrast to results for pristine samples of $\mathrm{TbTe}_3$ \citep{Maschek_prb_2015} and $\mathrm{DyTe}_3$ \citep{Maschek_prb_2018}. Our combined experimental results demonstrate the clear suppression of the secondary $a\text{-}\mathrm{CDW}$ phase at low intercalation levels and assign vestigial CDW features at $\boldsymbol{\mathrm{q}}_{1}^{a} \neq \boldsymbol{\mathrm{q}}_{2}^{a}$ in samples with $x \geq 0.02$ to effects of competing soft phonon modes at the onset of the $c\text{-}\mathrm{CDW}$ phase.

\section{ACKNOWLEDGMENTS}
This research used resources of the Advanced Photon Source, a U.S. Department of Energy (DOE) Office of Science user facility operated for the DOE Office of Science by Argonne National Laboratory under Contract No. DE-AC02-06CH11357. Crystal growth and characterization at Stanford University (JAWS, AGS, IRF) was supported by Department of Energy, Office of Basic Energy Sciences, under contract DE-AC02-76SF00515. Work by SR and RO at Argonne National Laboratory (XDS and IXS) was supported by the U.S. Department of Energy, Office of Science, Basic Energy Science, Materials Sciences and Engineering Division. RH acknowledges support by the state of Baden-Württemberg through bwHPC.

%

\end{document}